# Chiral skyrmions in an anisotropy gradient driven by spin-Hall effect

R. Tomasello[1], S. Komineas[2], G. Siracusano[3], M. Carpentieri[4], G. Finocchio[1]

[1] Department of Mathematical and Computer Sciences, Physical Sciences and Earth Sciences, University of Messina, I-98166, Messina, Italy

[2] Department of Mathematics and Applied Mathematics, University of Crete, 71003 Heraklion, Crete, Greece

[3] Department of Electric, Electronic and Computer Engineering, University of Catania, I-95125, Catania, Italy

[4] Department of Electrical and Information Engineering, Politecnico di Bari, via E. Orabona 4, I-70125 Bari, Italy

**Abstract**

A strategy to drive skyrmion motion by a combination of an anisotropy gradient and spin-Hall effect has recently been demonstrated. Here, we study the fundamental properties of this type of motion by combining micromagnetic simulations and a generalized Thiele's equation. We find that the anisotropy gradient drives the skyrmion mainly along the direction perpendicular to the gradient, due to the conservative part of the torque. There is some slower motion along the direction parallel to the anisotropy gradient due to damping torque. When an appropriate spin-Hall torque is added, the skyrmion velocity in the direction of the anisotropy gradient can be enhanced. This motion gives rise to acceleration of the skyrmion as this moves to regions of varying anisotropy. This phenomenon should be taken into account in experiments for the correct evaluation of the skyrmion velocity. We employ a Thiele-like formalism and derive expressions for the velocity and the acceleration of the skyrmion that match very well with micromagnetic simulation results.

# INTRODUCTION

Magnetic skyrmions are localized non-uniform magnetization patterns having topological features, i.e. they are characterized by an integer winding number $Q$ [1,2]. Such a non-trivial topology often provides an energy barrier which does not allow the skyrmion texture to be continuously wrapped to a spin configuration with a different $Q$. Skyrmions are becoming potential candidates to be used in low-power microelectronics applications, due to their attracting features, such as small size, energy stability, topological protection and manipulability via low spin-transfer torques, e.g. spin-Hall effect (SHE). The two basic ingredients to stabilize skyrmions are the use of materials with an out-of-plane easy-axis of the magnetization and the presence of a sufficiently large Dzyaloshinskii-Moriya interaction (DMI) [2–4]. According to the type of DMI (bulk or interfacial) and the material properties, different skyrmion chirality can be stabilized [2], i.e. Bloch, Néel and antiskyrmions [5]. Skyrmions were initially obtained in a non-linear field theory [6], and, more recently, chiral skyrmions (Bloch type) were predicted [7] and experimentally observed in a condensed matter system, namely in non-centrosymmetric cubic B20 materials [8–10]. On the other hand, Néel skyrmions have been experimentally stabilized at room temperature in multilayers systems [11–18], where the presence of ferromagnet/heavy metal interfaces gives rise to the interfacial DMI (IDMI). In this work, we will focus on Néel skyrmions (hedgehog-like configurations, see Fig. 1(a)), but our results can be also extended to Bloch skyrmions.

On the technological side, the possibility to electrically manipulate skyrmions [19–24] (nucleation, shifting, and detection) by means of spin-polarized currents (either spin-transfer torque [25] or SHE [26,27]), open the way for many promising applications as information carriers in low-power microelectronic technologies [20,28–31]. Recently, the single skyrmion nucleation (information writing) has been experimentally achieved [18], therefore now the efforts should be directed towards the analysis of the current-driven skyrmion motion. In particular, this motion is characterized by an in-plane angle with respect to the direction of the applied current, i.e. the skyrmion Hall angle. In an ideal system, micromagnetic simulations and theoretical approaches based on Thiele's equation [21,32] predict a constant skyrmion Hall angle independent of the value of the applied current. On the contrary, recent experimental observations [15,16] have shown a current dependence of the skyrmion Hall angle. This dependence can be explained by considering skyrmion-defects interaction [33–35]. In more details, a threshold current has to be overcome in order to move skyrmions in the presence of pinning and the skyrmion motion exhibits two dynamical regimes: for low currents, the skyrmion Hall angle is current dependent while, for larger currents, skyrmions are driven along a well-defined trajectory yielding a constant skyrmion Hall angle. In the latter regime, the results of micromagnetic simulations in ideal sample (no defects) are recovered. Recently, Yu *et al.* [17] have studied the skyrmion motion driven by SHE in the presence of a perpendicular anisotropy linear gradient. Our study is motivated by that work and we show, by means of micromagnetic simulations, that (*i*) the presence of only the anisotropy gradient induces a skyrmion motion with a main component along the direction perpendicular to the gradient itself, and (*ii*) the skyrmion is accelerated when the spin-polarization of the SHE is parallel to the gradient direction. In addition, we have generalized Thiele's equation [21,32] for the skyrmion velocity, to take into account the effect of the perpendicular anisotropy linear gradient, and we have derived an expression for the skyrmion acceleration under the hypothesis of adiabatic change of the skyrmion profile during its motion. We stress that in this paper we will consider an ideal sample hence the effect of the disorder is not considered.

The paper is organized as follows. Section 2 describes the micromagnetic framework as well as the derivation of the generalized Thiele's equation. Section 3 is devoted to the discussion of the analytical and micromagnetic results, while Section 4 shows the conclusions.

## MICROMAGNETIC MODEL AND THIELE'S EQUATION DERIVATION

We consider a ferromagnet in a square geometry with side lengths $L_x = L_y = 400 \text{ nm}$ and a thickness $t_{FM} = 1 \text{ nm}$. This is coupled to a layer of a heavy metal (Platinum) in order to achieve a sufficiently large IDMI (see Fig. 1(b)). Statics and dynamics of the magnetization are given by the Landau-Lifshitz-Gilbert equation that includes a Slonczewski term to take into account the spin-torque effect deriving from the SHE:

$$\partial_\tau \mathbf{m} = -\mathbf{m} \times \mathbf{h}_{eff} + \alpha \mathbf{m} \times \partial_\tau \mathbf{m} - b \mathbf{m} \times (\mathbf{m} \times \hat{\mathbf{e}}_y), \qquad (1)$$

where $\mathbf{m} = \mathbf{M}/M_s$ is the normalized magnetization, with $M_s$ being the saturation magnetization. The time variable $t$ has been redefined as $\tau = t/\tau_0, \tau_0 = 1/(\gamma_0 M_s)$, with $\gamma_0$ being the gyromagnetic ratio. The dimensionless spin-torque parameter is given by

$$b = \frac{j_{HM}}{j_0}, \qquad j_0 = \frac{2e}{\hbar} \mu_0 M_s^2 t_{FM} \frac{1}{\theta_{SH}} \qquad (2)$$

where $j_{HM}$ is the in-plane current injected via the heavy metal along the $x$-direction (see Fig. 1(a)), $e$ is the electron change, $\hbar$ is the reduced Planck's constant, $\mu_0$ is the permeability of the vacuum, $t_{FM}$ is the thickness of the ferromagnetic layer and $\theta_{SH} = 0.10$ is the spin-Hall angle [36]. We consider a spin polarization in the $y$-axis with unit vector $\hat{\mathbf{e}}_y$.

The effective field has the following expression

$$\mathbf{h}_{eff} = \Delta \mathbf{m} + \kappa m_z \hat{\mathbf{e}}_z + 2\lambda \left( \nabla m_z - \nabla \mathbf{m} \hat{\mathbf{e}}_z \right) + \mathbf{h}_m + \mathbf{h}_{ext} \qquad (3)$$

where we have included the exchange, perpendicular anisotropy, IDMI, magnetostatic terms and external fields, while $\hat{\mathbf{e}}_z$ is the unit normal to the film plane. Eqs. (1) and (3) are expressed in dimensionless form (the spatial dimensions are in exchange length unit). In particular, we consider the magnetic fields normalized to the saturation magnetization $M_s$. By considering that $A$ is the exchange constant, $K_u$ is the perpendicular anisotropy constant and $D$ is the IDMI parameter, we can introduce the exchange length $l_{ex} = \sqrt{2A/(\mu_0 M_s^2)}$, the characteristic length in presence of DMI $l_D = 2A/D$, the dimensionless anisotropy parameter $\kappa = 2K_u/(\mu_0 M_s^2)$, and the dimensionless DMI parameter $\lambda = l_{ex}/l_D$. In this work, we use the parameter values $M_s$=$10^6$ A/m, $A$=20 pJ/m, $D$=2.0 mJ/m$^2$, $K_u$=0.80 MJ/m$^3$, Gilbert damping constant $\alpha$=0.03, and then we obtain $l_{ex} = 5.64$ nm, $l_D = 20$ nm, $\kappa = 1.27$, $\lambda = 0.28$, and $j_0 = 1.91 \times 10^{13}$ A/m$^2$.

The micromagnetic computations are carried out by means of a state-of-the-art micromagnetic solver which integrates numerically Eq. (1) by applying the time solver scheme Adams-Bashforth [37], and the post-process is performed by state-of-the-art numerical tools [38]. The IDMI boundary conditions of the ferromagnetic sample are applied in the following way: $\frac{d\mathbf{m}}{dn} = \frac{D}{2A} (\hat{\mathbf{e}}_z \times \mathbf{n}) \times \mathbf{m}$ where $\mathbf{n}$ is the unit vector normal to the edge [39]. The used discretization cell is $2 \times 2 \times 1$ nm$^3$ (see Ref. [40] for a detailed description of the numerical micromagnetic framework). We discuss in detail micromagnetic simulations considering the aforementioned physical parameters [21], but qualitatively similar results have been obtained for a range of parameters. An out-of-plane field $H_{ext}$=50 mT ($h$=$H_{ext}/\mu_0 M_s$ =0.04) is applied in order to keep the skyrmion size relatively small and to reduce the transient breathing mode due to the applied current [21].

For the derivation of Thiele's equation, we assume the ultrathin film approximation for the magnetostatic field $\mathbf{h}_m = -m_z\hat{\mathbf{e}}_z$, hence we have an effective anisotropy constant $\kappa_{\text{eff}} = \kappa - 1 = 0.27$ .

We consider devices fabricated such that a constant linear anisotropy gradient $G$ is present along the $y$-direction of the sample [17]. In both the 3-dimensional sketch (Fig. 1(b)) and the top view (Fig. 1(c)) of the sample under investigation, the direction of $G$ is represented by a black arrow. Fig. 1(c) also displays the maximum ($K_{u,max}$) and minimum ($K_{u,min}$) values of the anisotropy constant $K_u$ at the edges of the sample. Table 1 shows the values of $K_{u,max}$ and $K_{u,min}$ as well as the corresponding value of $G$. At the center of the sample, the anisotropy coefficient is $K_u$=0.80 MJ/m$^3$.

| $K_{u,max}$(MJ/m$^3$) | $K_{u,min}$(MJ/m$^3$) | $G$(TJ/m$^4$) |
|---|---|---|
| 0.84 | 0.76 | 0.2 |
| 0.86 | 0.74 | 0.3 |
| 0.88 | 0.72 | 0.4 |
| 0.90 | 0.70 | 0.5 |
| 0.92 | 0.68 | 0.6 |
| 0.94 | 0.66 | 0.7 |
| 0.96 | 0.64 | 0.8 |
| 0.98 | 0.62 | 0.9 |
| 1.00 | 0.60 | 1.0 |

Table 1: Value of the linear gradient $G$ corresponding to the maximum and minimum $K_u$ values set at the sample edges along the $y$-direction.

The anisotropy coefficient is written as $K = K_u + Gy$, while the dimensionless anisotropy coefficient is given by

$$\kappa + gy \quad \text{where} \quad g = \frac{G}{G_0}, \quad G_0 = \frac{\mu_0 M_s^2}{2 l_{\text{ex}}}, \tag{4}$$

where $y$ is now understood to be dimensionless and in unit of $l_{\text{ex}}$. For our parameter values, the anisotropy gradient unit has the value $G_0 = 1.11 \times 10^{14}$ J/m$^4$.

The dynamical behavior of the skyrmion is probed by both the forces due to the anisotropy gradient and the spin-Hall torques. In order to derive an equation for the skyrmion dynamics, we assume that the skyrmion travels with a constant velocity $v = (v_x, v_y)$ and write

$$\mathbf{m} = \mathbf{m}(x - v_x t, y - v_y t), \quad \text{thus} \quad \partial_t \mathbf{m} = -v_k \partial_k \mathbf{m}, \tag{5}$$

where we use the notation $\partial_1 = \partial_x, \partial_2 = \partial_y$. We substitute the latter in Eq. (1) to obtain

$$v_\kappa \partial_\kappa \mathbf{m} = \mathbf{m} \times \mathbf{f} + \alpha v_\kappa \mathbf{m} \times \partial_\kappa \mathbf{m} + b \mathbf{m} \times (\mathbf{m} \times \hat{\mathbf{e}}_y),$$

then take the external product from the right with $\partial_\mu \mathbf{m}$ and subsequently the scalar product with $\mathbf{m}$:

$$v_\kappa \varepsilon_{\mu\kappa} q + \alpha v_\kappa \partial_\mu \mathbf{m} \cdot \partial_\kappa \mathbf{m} = -\mathbf{f} \cdot \partial_\mu \mathbf{m} + b(\mathbf{m} \times \partial_\mu \mathbf{m}) \cdot \hat{\mathbf{e}}_y$$

We will take advantage of the result that the integral over the $xy$-plane of the term $\mathbf{f} \cdot \partial_\mu \mathbf{m}$ vanishes for all contributions in $\mathbf{f}$ that derive from an energy functional that is invariant with respect to space translations [41]. We integrate over the $xy$-plane and obtain the system of equations for the velocity components

$$\begin{aligned} \alpha d_{11} v_1 + (Q + \alpha d_{12}) v_2 &= -C_1 - bT_1 \\ (-Q + \alpha d_{21}) v_1 + \alpha d_{22} v_2 &= -C_2 - bT_2 \end{aligned} \tag{6}$$

where $Q$ is the skyrmion number and other quantities are defined as

$$\begin{aligned} C_\mu &= \frac{1}{4\pi} \int (\mathbf{f}_{\text{ext}} \cdot \partial_\mu \mathbf{m}) dx dy \\ T_\mu &= \frac{1}{4\pi} \int (\partial_\mu \mathbf{m} \times \mathbf{m}) \cdot \hat{\mathbf{e}}_y dx dy \\ d_{\mu\nu} &= \frac{1}{4\pi} \int (\partial_\mu \mathbf{m} \cdot \partial_\nu \mathbf{m}) dx dy, \end{aligned} \tag{7}$$

with $\mu = 1, 2$ denoting the $x, y$ directions. Note that the quantities $C_\mu$ have contributions only from forces deriving from a space dependent field and this is here the anisotropy gradient $\mathbf{f}_{\text{ext}} = gy m_z \hat{\mathbf{e}}_z$. The quantities, $C_\mu$, $T_\mu$ and $d_{\mu\nu}$ depend on the skyrmion profile and they are time dependent since the skyrmion profile is expected to change as the skyrmion is moving. This is indeed what we observe in our system and it will be discussed in detail.

The quantities $C_\mu$ give contributions to the dynamical equations (6), and they are $C_1 = 0$ and

$$C_2 = \frac{1}{4\pi}\int g y m_z \partial_y m_z dx dy = \frac{g}{8\pi}\int (1-m_z^2)dx dy,$$

where we have applied a partial integration in the last step. We write the above as

$$C_1 = 0, \quad C_2 = g\chi, \qquad \chi \equiv \frac{1}{8\pi}\int(1-m_z^2)dx dy. \qquad (8)$$

The spin-torque gives contributions in the dynamical equations (6) through $T_1$, $T_2$. The expressions are simplified if we assume that the skyrmion profile remains axially symmetric and Néel-type throughout the motion. We write this axially symmetric profile using the angle variables for the spherical parameterization [21]

$$\Theta = \Theta(\rho), \qquad \Phi = \phi \qquad (9)$$

where $\rho, \phi$ are polar coordinates, and we find

$$T_1 = T, \; T_2 = 0, \; T \equiv \frac{1}{8\pi}\int\left(\partial_\rho\Theta + \frac{\sin(2\Theta)}{2\rho}\right)(2\pi\rho d\rho) \qquad (10)$$

We use the simplified expressions of Eqs. (8) and (10) in the dynamical equations (6) and obtain the velocity

$$v_x = g\chi - \alpha d(bT), \qquad v_y = -bT - ad(g\chi) \qquad (11)$$

where we have omitted terms $O(\alpha^2)$ and we have set $Q = 1$. We have further defined $d \equiv d_{11} = d_{22}$ and set $d_{12} = d_{21} = 0$ which is correct for an axially symmetric configuration and it is a very good approximation as verified in our simulations. In order to obtain the velocity of Eq. (11) in SI units, one should multiply by $\upsilon_0 = \ell_{ex}/\tau_0 = 1250$ m/s.

Eqs. (11) is valid for a skyrmion moving with constant velocity. However, a constant velocity can be obtained only when the skyrmion profile is space independent. In our simulations, the anisotropy parameter depends on the position and the skyrmion profile is changing as this is moving. In Eq. (11), $\chi$ is proportional to the anisotropy energy, while $T$ is proportional to the absolute value of the IDM energy, that means that the velocity of the skyrmion depends on its position in a sample where the anisotropy varies. In regions of lower anisotropy, the skyrmion would expand [30] and thus have larger values for the integral $\chi$ in Eq.

(8), and larger values for the integral $T$ in Eq. (10). The quantity $d$ is scale invariant, i.e. simple variations of the skyrmion radius would leave $d$ invariant. The quantity $d$ would only change when the form of the skyrmion profile would change, but such effects are very small. Neglecting the effect of damping in Eq. (11), we expect that $v_x$ would be larger in regions of negative $y$ for the sample in Fig. 1(b) due to the anisotropy gradient, and $v_y$ would be larger in the same regions due to the SHE torque. In the following, we will use these considerations in order to analyse the numerical results.

The velocity variations due to changes of the anisotropy parameter and, subsequently of the $\chi$ and $T$ values, call for the definition of a skyrmion acceleration. In this case, Eq. (11) could be still valid as an approximation for the *instantaneous* velocity if we assume that the skyrmion adjusts its profile (and diameter) *adiabatically* to the local anisotropy as it moves. We then apply Eq. (11) for values of $\chi$ and $T$ that are calculated for a static skyrmion which would be subject to the local anisotropy at the instantaneous position of the moving skyrmion.

In our study, the anisotropy constant varies only in the $y$-direction, therefore $\chi = \chi(y)$, and $T = T(y)$. By taking the time derivative of Eq. (11), the acceleration is obtained:

$$a_x = -\left[ bT + (\alpha d)(g\chi) \right]\left[ g\chi' - (\alpha d)bT' \right]$$
$$a_y = \left[ bT + (\alpha d)(g\chi) \right]\left[ bT' + (\alpha d)g\chi' \right]$$

(12)

where $\chi' = \dfrac{d\chi}{dy}$ and $T' = \dfrac{dT}{dy}$. We have neglected the variation of $d$ in the calculation of the acceleration because this is very small, as confirmed in our numerical calculations. For the parameters in our simulations, and in related experiments, we have that $\alpha$, $g$, $b$ are all small (much smaller than unity) and they take similar values. We can also assume that both quantities $\chi$ and $T$ vary linearly with $\kappa$ to a first approximation therefore $\chi'$ and $T'$ are $O(g)$ and that $d$ is constant. These assumptions are verified by numerical calculations. In this case, the lowest order terms for the acceleration in Eq. (12) are:

$$a_x = -bgT\chi', \ a_y = b^2 TT'.$$

(13)

## RESULTS AND DISCUSSION

For our set of parameters, the equilibrium skyrmion diameter is 20 nm. It is calculated as the diameter of the skyrmion core defined as the region where the $z$-component of the magnetization is $m_z < 0$. The position of the skyrmion is iteratively tracked by identifying its core coordinates from the skyrmion shape (see supplementary information in Ref. [42]). This provides an accurate and reliable methodology to determine the time evolution of its trajectory. Once we have evaluated the position at successive time instants $t_i$, we then calculate the velocity from the difference of successive positions, while the acceleration is found from the fitting of the position as a function of time by a parabola.

First, we analyze the well-known [21] skyrmion motion driven by the SHE only (see Figs. 2(a) and (b)). As expected from Eq. (11), the skyrmion exhibits a major velocity component $v_y$ along the negative $y$-axis as well as a smaller $x$-component $v_x$ in the negative $x$-direction. The micromagnetic results are in good agreement with the analytical ones obtained via Thiele's equation.

Figs. 2(c) and (d) show the velocity components when the skyrmion is driven by an anisotropy gradient only. The range of $G$ has been chosen such that the anisotropy field difference across our 20 nm skyrmion is similar to the experimental value [17] (obtained for a field gradient 9 MJ/m$^4$ acting on skyrmions with a diameter of thousands of nanometers). Both velocity components increase with the gradient and the agreement between analytical results and micromagnetics is good. In this case, the skyrmion is characterized by a larger $v_x$ in the positive $x$-direction and a very small negative $v_y$ (see Movie 1). For the computation of the analytical velocities by Eq. (11), we have used the initial profile of the skyrmion in the center of the sample, where $K_u = 0.80$ MJ/m$^3$. This should be a good approximation since the displacement in the $y$-direction is much smaller than the one in the $x$-direction, and thus the skyrmion expansion can be neglected. The slight discrepancy between analytics and micromagnetics can be mainly ascribed to this approximation.

The skyrmion motion induced by the anisotropy gradient can be understood as follows (see inset in Fig. 2(c)). The presence of the anisotropy gradient gives rise to different values of the effective field for the upper and lower in-plane magnetization of the skyrmion domain wall, where the magnetization is mainly oriented along the $y$-direction, thus generating two different conservative torques acting on the skyrmion. These torques do not balance each other leading the skyrmion to move along the $x$-direction. According to the direction of $G$ and/or the skyrmion polarity, the skyrmion can travel along the positive $x$-direction, as in the case analysed here where the upper conservative torque is larger, or, vice versa, along the negative $x$-direction if $G$ and/or the polarity are reversed. The motion along the $y$-direction is due to the damping torque (see Eq. (11)). This is also confirmed by our micromagnetic simulations where no $v_y$ is observed when we set to zero the damping term in Eq. (1) (not shown).

Based on our theoretical results, one would expect to experimentally observe a shift of skyrmions on one side of the sample. However, such a behavior has not been observed in experiments [17]. This is due to the presence of defects in real sample which play a fundamental role as pinning centers for the skyrmion [33–35,43].

In order to study the combined effect of the anisotropy gradient and SHE, we perform a simulation by fixing $G$=0.5 TJ/m$^4$ and $j_{HM}$=10 MA/cm$^2$. The skyrmion moves to the positive $x$-direction mainly due to the anisotropy gradient and to the negative $y$-direction mainly due to the SHE torque, as it is expected by Eq. (11). There is a negative contribution to $v_x$ due to damping and SHE, but this is completely compensated by the positive contribution of the conservative term due to the anisotropy gradient. The component $v_y$ has a negative contribution due to both terms in Eq. (11). As the skyrmion moves to regions of smaller anisotropy (negative $y$), we observe an accelerated motion of the skyrmion in both $x$ and $y$ directions, due to the skyrmion expansion (see Movie 2). This is a key result of this work. In order to make a quantitative comparison between analytics and micromagnetics for benchmarking the generalized Thiele's equation, we integrate Eq. (11) over time to obtain the analytical skyrmion position as a function of time for both $x$- and $y$-directions, and then we compare these data with the micromagnetic results (see Figs. 3(a) and (b)). We observe that both trajectories show good agreement and, when compared with a parabolic fit, we can conclude that a uniform accelerated motion of the skyrmion is achieved for short current pulses. From those data, we can extract the accelerations $a_x$ and $a_y$ as half of the coefficient of the squared term.

Figs. 4(a) and (b) show the two components of the acceleration as functions of current, where the analytical results are in very good agreement with the micromagnetic outcomes. The increase of $|a_y|$ with current is expected since the SHE-driven skyrmion motion is mainly along the negative $y$-direction [21] (see Fig. 1(b)). The motion in the positive $x$-direction is mainly due to the gradient but this is partially compensated by the SHE that gives a negative contribution. Thus, the acceleration component $a_x$ increases with current more moderately. In addition, those figures show the accelerations as computed directly via Eq. (13) by using the initial profile of the skyrmion in the center of the sample, where $K_u$=0.80 MJ/m$^3$. The agreement is still very good for the major component $a_y$, while $a_x$ is somewhat overestimated. Nonetheless, we believe that Eq. (13) can be used for a first simple estimation of the acceleration.

By means of the generalized Thiele's equation, we can calculate the instantaneous skyrmion velocity as a function of the local perpendicular anisotropy in correspondence of the position of the skyrmion. As already discussed in Section 2, the observed increase of the skyrmion velocity can be anticipated by Eq. (11) if we assume that the skyrmion changes its diameter [30,44] as it moves to the negative $y$-direction (lower anisotropy region) and adiabatically adjusts its profile to the local anisotropy strength. We plot the results in Fig. 4(c) and (d) where we find a very good agreement between analytics and micromagnetics for both velocity components.

## SUMMARY AND CONCLUSIONS

In summary, we have studied the spin-Hall effect-driven skyrmion motion in presence of a perpendicular anisotropy linear gradient by means of micromagnetic simulations and a generalized Thiele's equation that includes the linear anisotropy gradient. We have observed that the anisotropy gradient introduces a major component of the skyrmion velocity in the direction perpendicular to the gradient. Moreover, when the skyrmion moves towards a region with lower perpendicular anisotropy, its diameter increases, thus inducing an accelerated motion. We have derived expressions for the skyrmion velocity and acceleration from the generalized Thiele's equation (Eq. (11) and (13)) by considering an adiabatic change of the skyrmion size that agrees well with results of full micromagnetic simulations. Our results can drive the design of skyrmion based devices for storage (e.g. racetrack memories), logic [45], and unconventional computing [46], that take the advantage of anisotropy gradients combined with SHE.

## ACKNOWLEDGMENTS

This work was supported by the executive program of scientific and technological cooperation between Italy and China for the years 2016-2018 (code CN16GR09) title "Nanoscale broadband spin-transfer-torque microwave detector" funded by Ministero degli Affari Esteri e della Cooperazione Internazionale, and by the bilateral agreement Italy-Turkey (TUBITAK-CNR) project (CNR Grant #: B52I14002910005, TUBITAK Grant # 113F378) "Nanoscale magnetic devices based on the coupling of Spintronics and Spinorbitronics".

**Figures' captions**

Figure 1: (a) The static axially symmetric ($Q=1$) Néel skyrmion with negative polarity and diameter of 20 nm represented through the spatial distribution of the magnetization as obtained from micromagnetic simulations (the colors are linked to the normalized $z$-component blue negative, red positive). (b) A schematic representation of the ferromagnet/heavy metal bilayer under investigation, where the current $j_{HM}$ and the gradient $G$ are also indicated. (c) Top view of the ferromagnetic layer. The skyrmion Hall angle, the directions of the gradient $G$ and current density in the heavy metal $j_{HM}$ are indicated.

Figure 2: (a) and (b) skyrmion velocity components ($v_x$ and $v_y$, respectively) as a function of $j_{HM}$ when only SHE acts. (c) and (d) skyrmion velocity components ($v_x$ and $v_y$, respectively) as a function of $G$ in presence of only the anisotropy gradient. In all the figures, the black circles represent the micromagnetic results, while the red line the analytical ones. Inset in (c): schematic representation of the skyrmion circular in-plane domain wall in the presence of the anisotropy gradient. The two different conservative torques are indicated.

Figure 3: Skyrmion position as a function of time along the (a) $x$-direction and (b) $y$-direction for $j_{HM}$=10 MA/cm$^2$ and $G$=0.5 TJ/m$^4$. The black circles represent the results as obtained by micromagnetic simulations, whereas the solid red line indicates the analytical results by solving numerically Eq. (11).

Figure 4: (a) and (b) Skyrmion acceleration components ($a_x$ and $a_y$, respectively) as functions of $j_{HM}$ for the SHE-driven motion in the presence of the anisotropy gradient ($G=0.5$T J/m$^4$), the black circles (red line) are computed by a fitting with a parabola of the micromagnetic data (analytical data), as explained for Fig. 3, while the blue dashed line is computed by using Eq. (13). (c) and (d) Skyrmion velocity components ($v_x$ and $v_y$, respectively) as functions of the perpendicular anisotropy constant along the gradient ($G=0.5$ J/m$^4$) when $j_{HM}$=10 MA/cm$^2$. In all entries of the figure, the black circles represent the micromagnetic results, while the red line the analytical ones.

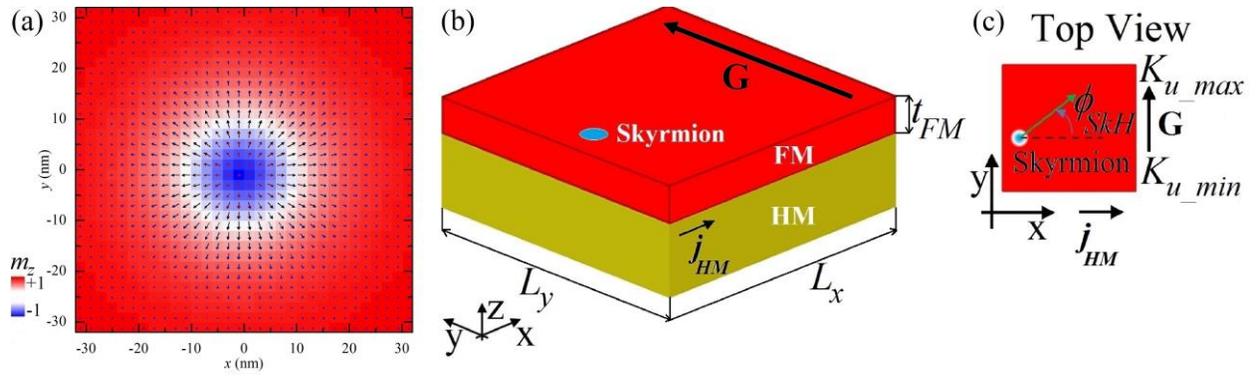

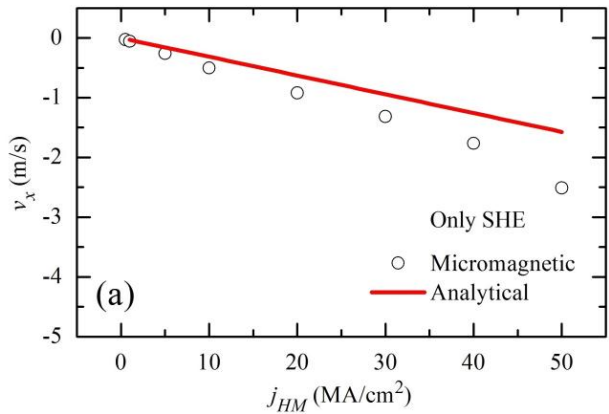

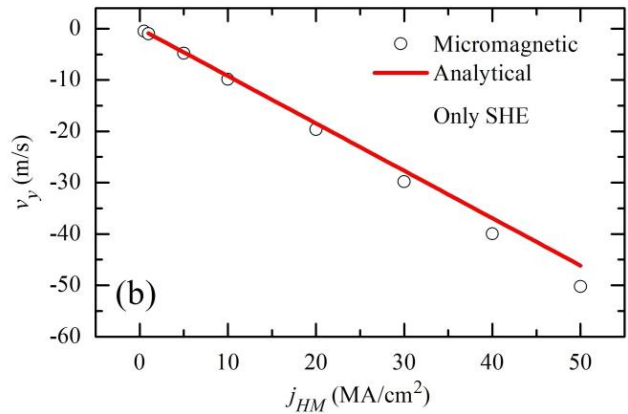

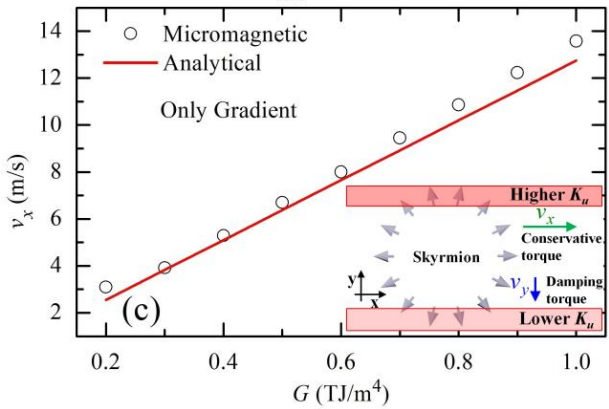

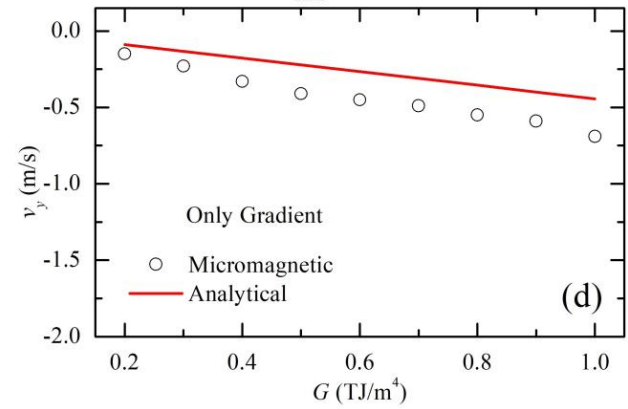

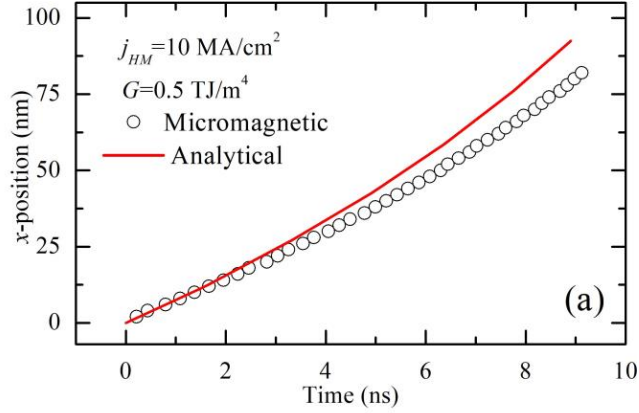

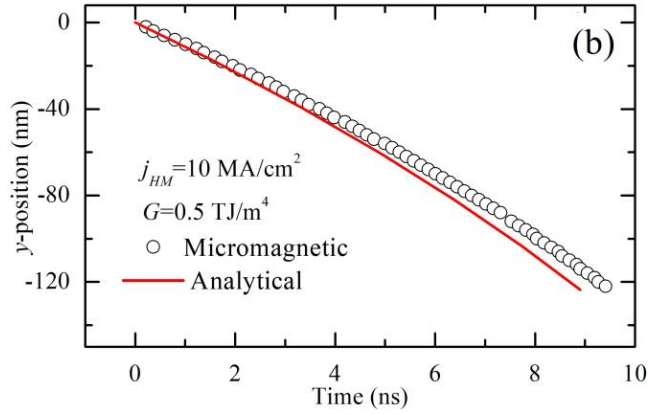

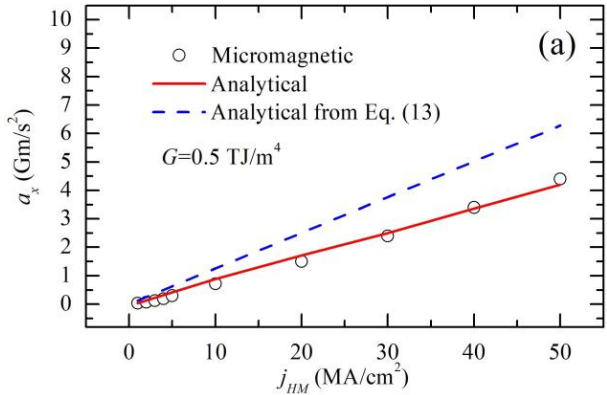

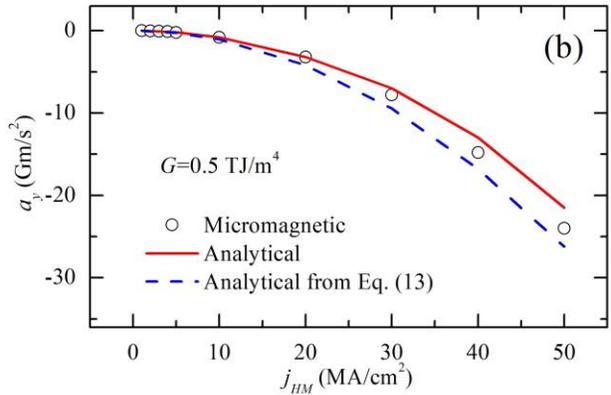

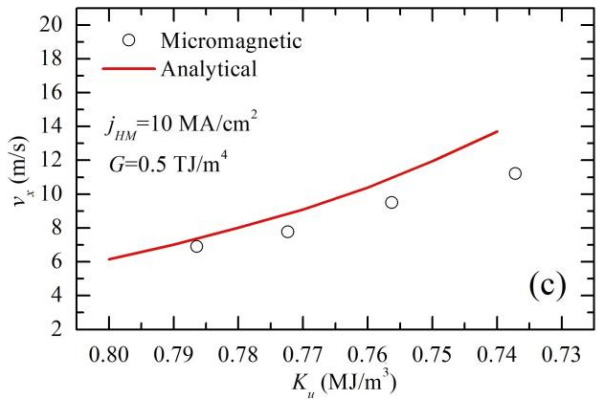

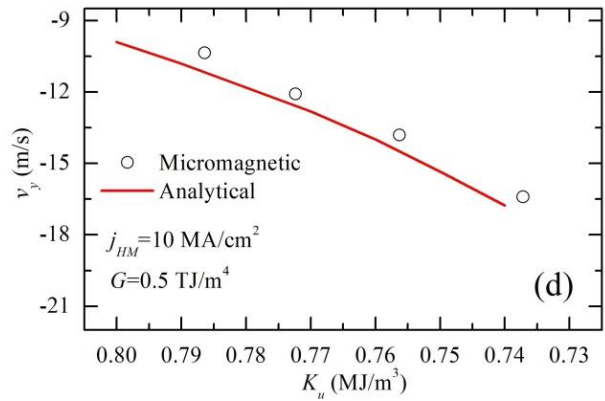